\documentclass{eusflat2021}

\usepackage{graphicx}
\usepackage{amssymb,amsfonts,amsthm,mathrsfs,mathtools}
\usepackage{float}
\usepackage{enumitem}
\usepackage{xcolor}
\usepackage[printonlyused]{acronym} 
\usepackage{blindtext}

\usepackage{physics}

\newcommand{\ie}{i.e.}

\newcommand{\scref}[1]{Section~\ref{#1}}

\newcommand{\fgref}[1]{Figure~\ref{#1}}
\newcommand{\tbref}[1]{Table~\ref{#1}}

\usepackage{bm}
\renewcommand{\vec}[1]{\mathbf{#1}}
\newcommand{\vecsym}[1]{\boldsymbol{#1}}

\DeclareMathOperator*{\argmax}{arg\,max}

\usepackage{tabularx} 
\usepackage{caption}
\usepackage{subfig}
\usepackage{multirow, makecell}
\usepackage{longtable}

\title{\bf Evolving Fuzzy System Applied to Battery Charge \\ Capacity Prediction for Fault Prognostics}
\author{Murilo Osorio Camargos$^a$ \and Iury Bessa$^{b,c}$ \and Luiz A. Q. Cordovil Junior$^b$\\ \and \textbf{Pedro Henrique Silva Coutinho}$^b$ \and \textbf{Daniel Furtado Leite}$^d$ \and $^*$\textbf{Reinaldo Martínez Palhares}$^e$\\
$^a$School of Applied Mathematics, Fundação Getulio Vargas, Rio de Janeiro, Brazil. \email{murilo.camargosf@gmail.com} \\
$^b$Graduate Program in Electrical Engineering, Federal University of Minas Gerais, Belo Horizonte, Brazil.\\\email{{luiz.cordovil05, coutinho.p92}@gmail.com} \\
$^c$Department of Electricity, Federal University of Amazonas, Manaus, Brazil. \email{iurybessa@ufam.edu.br}\\
$^d$Department of Automatics, Federal University of Lavras, Lavras, Brazil. \email{daniel.leite@ufla.br} \\
$^e$Department of Electronics Engineering, Federal University of Minas Gerais, Belo Horizonte, Brazil. \email{rpalhares@ufmg.br} }

\begin{document}

\maketitle

\begin{abstract}
This paper addresses the use of data-driven evolving techniques applied to fault prognostics. In such problems, accurate predictions of multiple steps ahead are essential for the \ac{rul} estimation of a given asset. The fault prognostics' solutions must be able to model the typical nonlinear behavior of the degradation processes of these assets, and be adaptable to each unit's particularities. In this context, the \acp{efs} are models capable of representing such behaviors, in addition of being able to deal with non-stationary behavior, also present in these problems. Moreover, a methodology to recursively track the model's estimation error is presented as a way to quantify uncertainties that are propagated in the long-term predictions. The well-established NASA's Li-ion batteries data set is used to evaluate the models. The experiments indicate that generic \acp{efs} can take advantage of both historical and stream data to estimate the \ac{rul} and its uncertainty.

\end{abstract}

\acresetall

\section*{Acronyms}
\begin{acronym}[xxxxxxxxx] 

    \acro{xts}[exTS]{Evolving Extended Takagi-Sugeno}
    \acro{rul}[RUL]{Remaining Useful Life}
    \acro{eol}[EOL]{End of Life}
    \acro{phm}[PHM]{Prognostics and Health Management}
    \acro{ebets}[EBeTS]{Error Based Evolving Takagi-Sugeno Model}
    \acro{cbm}[CBM]{Condition-based Maintenance}
    \acro{hi}[HI]{Health Index}
    \acro{ai}[AI]{Artificial Intelligence}
    \acro{pdf}[PDF]{Probability Density Function}
    \acro{ts}[TS]{Takagi-Sugeno}
    \acro{uut}[UUT]{Unit Under Test}
    \acro{fbem}[FBeM]{Fuzzy Set Based Evolving Modeling}
    \acro{mf}[MF]{Membership Function}
    \acro{rls}[RLS]{Recursive Least Squares}
    \acro{msr}[MSR]{Most Similar Rule}
    \acro{mape}[MAPE]{Mean Absolute Percentage Error}
    \acro{ra}[RA]{Relative Accuracy}
    \acro{lcr}[LCR]{Last Created Rule}
    \acro{rmse}[RMSE]{Root Mean Squared Error}
    \acro{rms}[RMS]{Root Mean Square}
    \acro{ann}[ANN]{Artificial Neural Network}
    \acro{ft}[FT]{Fault Threshold}
    \acro{pf}[PF]{Particle Filter}
    \acro{arma}[ARMA]{Autoregressive Moving Average}
    \acro{anfis}[ANFIS]{Adaptive Neuro-Fuzzy Inference System}
    \acro{nasa}[NASA]{National Aeronautics and Space Administration}
    \acro{ndei}[NDEI]{Non Dimensional Error Index}
    \acro{enfn}[eNFN]{Evolving Neo-Fuzzy Neural Network}
    \acro{emg}[eMG]{Evolving Multivariable Gaussian}
    \acro{lstm}[LSTM]{Long Short-Term Memory}
    \acro{efs}[EFS]{Evolving Fuzzy System}

    \acrodefplural{lmi}[LMIs]{Internets-of-Things}

\end{acronym}

\section{Introduction}


The industry has been investing many resources to create new maintenance policies that can prevent unexpected failures. These maintenance policies are in continuous improvement towards reliability and cost-effectiveness. Between preventive
and corrective maintenance strategies, \ac{cbm} is developed to be the optimal point in terms of total costs, balancing operating with maintenance costs.

A key \ac{cbm} program is the \ac{phm}, which creates relevant health indicators out of monitoring data to reduce inspections through early fault detection and 
prediction of impending faults~\cite{Jouin2016}. Health prognostics is a primary task in this context and consists of predicting the \ac{rul} of these 
machines~\cite{Lei2018}. 
The \ac{rul} prediction methodologies are commonly classified into three categories: data-driven, model-based, and hybrid. The latter combines characteristics of the former two categories \cite{Kan2015}. Model-based approaches rely on first-principle models to assess the RUL~\cite{Cubillo2016}.
Although these models tend to outperform models in other categories, they may be challenging to obtain in practical situations, and reusing these models in different assets may be impossible~\cite{Lei2018}. 

The drawbacks of first-principle methods motivate the development of data-driven approaches based on statistical models and artificial intelligence techniques.
%
%
Particularly, artificial intelligence approaches deal with complex systems by learning how to produce the desired outputs from given inputs, i.e, learning input-output relationships, which are possibly nonlinear \cite{An_2015}. However, it is usually necessary to retrain artificial intelligence models if the operating conditions change~\cite{Peng2010}; and the algorithms usually require a large amount of high-quality training data. 
	
In general, data-driven approaches have fixed structures, \ie, they assume stationary environment. Such assumption often does not hold, 
making the aforementioned approaches unsuitable for real-time prognostics, where human intervention is not always possible to redefine the problem domain if needed. 
A way of tackling the stationarity assumption is to develop strategies based on multiple models \cite{cosme2018novel}. The multiple model strategy can be automated by
developing evolving models, whose knowledge-base is built based on data streams, allowing the learning of complex behaviors and novelties from scratch~\cite{Cordovil2019}. 
The ability to model complex nonlinear dynamics in non-stationary environments places the \acp{efs} as interesting choices for prognostics applications in cases 
where it is rough to represent or describe time-varying and nonlinear characteristics of a system. 
Nonetheless, literature on applying evolving intelligence to fault prognostic issues is somewhat scarce; a subset of that addresses the uncertainty quantification 
problem \cite{Camargos2020,El-Koujok2011,Gouriveau2012,Ramasso2014}.

The practical applications of \acp{efs} are various. Their recursive nature allows real-time fault detection and diagnosis \cite{Leite2009,Lemos2013},
systems identification, and time-series prediction \cite{Leite2015}. The present study focuses on system identification and time-series prediction problems in which the majority of the existing models are based on variations of \ac{ts} fuzzy inference systems, \ie, models whose rules consist of functional consequent terms. 
In their evolving formulations, these models display a fully adaptive structure in terms of the number of rules, and antecedent and consequent parameters through data-streams.
Online learning is supported by a recursive incremental learning mechanism that decides about rule creation, exclusion, updating, and merging.

Throughout the years, different kinds of \acp{efs} have been proposed to explore nuances of different learning mechanisms. The multivariable model called \ac{xts} \cite{Angelov2006}
%
partitions the input/output data space through 
an extension of the concepts of subtractive clustering \cite{Chiu1994}. The learning process of creating and excluding rules in 
the knowledge-base is related to recursively computed quality metrics such as the zone of influence of each cluster, their age, and support size. 
Local models are constructed by means of univariate Gaussian \acp{mf} for each premise variable.

The application of evolving models for time series prediction and systems identification has stimulated efforts towards the development of models that account for complex relationships among input variables. The \ac{efs} called \ac{emg} uses first-order functions and multivariate Gaussians 
as \ac{mf} to represent the premise variables \cite{Lemos2011}. This kind of \ac{mf} can model the relation among input variables through the recursive computation of a dispersion matrix. The model uses a learning mechanism based on the participatory learning principle that endows the algorithm with the 
capacity to classify whether a sample is an outlier or the first representative of a new cluster \cite{Yager1990}.

Due to a clear relation between the prognostics task addressed in this paper and long-term forecasting, the \ac{ebets} is considered. \ac{ebets} combines multivariate Gaussians -- to represent complex relationships among input variables -- with criteria designed to explicitly take advantage of the estimation error to update the model's structure on the fly \cite{Camargos2020}. The choice of hyper-parameters in \ac{ebets} can be made in a fully problem-agnostic way, which facilitates its application in different problems, such as the prognostics of rolling bearing and Li-ion batteries. Reference \cite{Camargos2020} also provides a framework that enables 
the use of different \acp{efs} in prognostics tasks in which model's uncertainty must be considered. The main contributions of the present paper are the following:
\begin{enumerate}
    \item Fault prognostics is performed taking into consideration the Li-ion battery dataset and \acp{efs};
    \item The uncertainty quantification procedure proposed in \cite{Camargos2020} is improved by means of a more stable quantification of the model's initial error;
    \item The \ac{rul}'s confidence bounds were generalized as z-values of the normal distribution computed through a given significance level.
\end{enumerate}

The remainder of this paper is organized as follows. \scref{sec:prog_prob} states the prognostics problem by providing the definition of \ac{rul} used in this paper. \scref{sec:ddp_efs} describes the computational framework that supports \acp{efs} for long-term prediction as well as the novel uncertainty quantification procedure, which allows the computation of confidence intervals using fuzzy \ac{ts} models. In \scref{sec:exp_set}, a real benchmark dataset, namely, the Li-ion battery charging dataset, provided by NASA, and a parameter tuning procedure to test prognostics approaches are presented. \scref{sec:res_dis} shows the results and discusses the application of three different \acp{efs} and a non-evolving model for fault prognostics. Finally, \scref{sec:conc} concludes the paper.

\section{The prognostics problem}
\label{sec:prog_prob}

\ac{rul} prediction is an essential step in prognostics. Based on the asset's age and condition, and on previous operation profile, \ac{rul} prediction concerns estimating how much time remains, from the current instant, to a possible
fault occurrence \cite{Jardine2006}. Some authors define \ac{rul} from the \ac{hi} 
point of view, \ie, \ac{rul} is the time left until the system's degradation 
state reaches a given \ac{ft} \cite{Li2015,Si2013}, which is expressed by:
\begin{equation}
	\hat{r}_k = \inf{\left\{N\in\mathbb{N} \,\vert\, \hat{x}_{k+N} \geq \eta\right\}},
	\label{eq:rul_def}
\end{equation}
where $\hat{r}_k$ denotes the \ac{rul} computed at instant $k$, given observations of the degradation state until $k$; $\mathbb{N}$ is the natural numbers set; $\hat{x}_{k+N}$ is an estimate of the degradation state, \ie, the \ac{hi} at time $k+N$; and $\eta$ is the predefined \ac{ft}.

\begin{figure}[h]
	\centering
	\includegraphics[width=\linewidth]{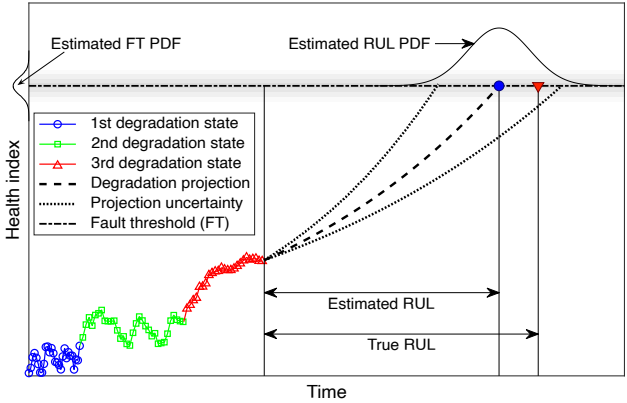}
	\caption{Degradation stages and uncertainty in \acs{rul} prediction.}
	\label{fig:ch2_prognostics}
\end{figure}

In addition to a pointwise \ac{rul} estimate, providing a confidence interval in which the \ac{rul} belongs -- which takes into consideration the inherent uncertainty of fault prognostics \cite{TobonMejia2012TR} -- is equally important. An example of uncertainty in \ac{rul} estimate is shown in \fgref{fig:ch2_prognostics}. Notice that the predicted degradation path reaches the \ac{ft} (blue dot) before the actual degradation path does (red dot). However, a confidence interval extracted from a probability density function around the pointwise estimate encloses the true \ac{rul}. Another uncertain value in \ac{rul} prediction is related to the \ac{ft} itself. The \ac{ft} can also be described by means of a probability distribution -- or \textit{failure domain}. However, in a great part of the literature, including the present paper, the \ac{ft} is represented by a constant line to simplify the \ac{rul} prediction process \cite{Lei2018}. Therefore, Eq. \eqref{eq:rul_def} is a simplified version of the canonical \ac{rul} definition by \cite{Chiachio2015}, where the concept of failure domain is defined.

The state of a system deteriorates until it reaches the \ac{ft}, namely $\eta$. Thus, we define the \ac{rul} prediction problem in this paper as a multi-step-ahead prediction problem; we want to estimate $\eta$. State propagation can be performed using the following state transition relation,
\begin{equation}
	\hat{x}_{k+N|k} = f_k\left(\vec{v}_{k+N,L}, \varepsilon_{k+N}\right),
	\label{eq:state_transition}
\end{equation}
in which $\vec{v}_{k+N,L}$, given by
\begin{equation}
    \small
    \vec{v}_{k+N,L}^\top = \begin{cases} \left[x_{k} \quad x_{k-1} \quad
    \cdots \quad 
    x_{k-L+1}\right], & \text{if } N = 1\\
    \left[\hat{x}_{k+N-1}\enskip\cdots\enskip\hat{x}_{k+1}\enskip
    x_{k}\enskip\cdots\enskip x_{k+N-L}\right], & \text{if } 2\leq N \leq L \\
    \left[\hat{x}_{k+N-1}\quad\cdots 
    \quad\hat{x}_{k+N-L}\right], & 
    \text{if } N > L
    \end{cases}
    \label{eq:xkl_lags_nmdl}
\end{equation}
is a lag vector with estimates; $\varepsilon_k$ is an independent identically distributed (i.i.d.) noise vector. Furthermore, $x_n$ and $\hat{x}_n$ are, respectively, the observed and estimated degradation state at the time step $n$; $L$ is the order of the auto-regression polynomial; $N$ is the number of steps ahead for which the degradation state is predicted; and $f_k(\cdot)$ is the state transition function recursively obtained up to the instant $k$ using an \ac{efs}. In this paper, long-term estimates are given by the iterative approach \cite{Gouriveau2012} due to its simple implementation and quickness. Moreover, establishing a prediction horizon is needless. The iterative approach performs one-step prediction, and uses the last predicted value as a regressor to estimate the next value.

\section{Data-driven prognostics with \ac{efs}}
\label{sec:ddp_efs}

A \ac{ts} fuzzy model allows the representation of a system by means of fuzzy concepts. The \ac{ts} fuzzy model uses functional 
consequent, usually linear \cite{Fuzzy-Past-Present-Future2019}. In such 
systems, given a set of $C$ rules, the $i$-th \textbf{IF-THEN} rule, is
\begin{equation}
\begin{aligned}
\mathrm{Rule}\,\,i\mbox{: } & \textsc{if } \vec{x}_k \textsc{ is } \Phi_{i,k-1} \textsc{ then } \hat{y}_{i,k} = \tilde{\vec{x}}_k^\top \, \hat{\vb*{\theta}}_{i,k-1}
\end{aligned}
\label{eq:ith_fuzzy_rule}
\end{equation}
where $\vec{x}_{k} \in \mathbb{R}^{n_x}$ is the vector of premise variables, $\vecsym{\hat{\theta}}_{i,k-1}\in\mathbb{R}^{n_x+1}$ is the vector of estimated consequent parameters, and $\tilde{\vec{x}}_k = \left[1 \quad \vec{x}_k^\top\right]^\top$. Moreover, $\vec{x}_k \textsc{ is } \Phi_{i,k-1}$ denotes the fuzzy relation between $\vec{x}_k$ and the fuzzy set $\Phi_{i,k-1}$ for $i\in\mathbb{N}_{\leq C}$, such that the multivariate \ac{mf} is $\varphi_{i,k-1}\colon\mathbb{R}^{n_x}\to[0,1]$. Throughout the text, $\mathbb{N}_{\leq k}$ will be used to denote the set of natural numbers up to $k$, such that $\mathbb{N}_{\leq k}=\{1,2,\dots,k\}$.

The output of the \ac{ts} fuzzy model is a convex combination among $C$ consequent linear models weighted by the rules' activation degrees. Activation degrees must comply with the convex sum property, \ie, they need to be non-negative and sum one. From the center average defuzzification, the overall model output is given as
\begin{equation}
\hat{y}_k = \sum_{i=1}^C h_{i,k-1}(\vec{x}_k) \,\hat{y}_{i,k}
\label{eq:ts_model_output}
\end{equation}
in which
\begin{equation}
h_{i,k-1}(\vec{x}_k) = \frac{\varphi_{i,k-1}(\vec{x}_k)}{\sum_{m=1}^C \varphi_{m,k-1}(\vec{x}_k)}.
\label{eq:ts_model_weights}
\end{equation}

When using univariate \acp{mf}, rules' activation degrees are obtained from an implication operator (aggregation function), quite often a t-norm. At instant $k$, the system output \eqref{eq:ts_model_output} can be rewritten in matrix form as
\begin{equation}
    \hat{y}_k = \vec{h}_{k-1}^\top(\vec{x}_k) \,\vecsym{\hat{\Theta}}_{k-1}^\top 
    \, \tilde{\vec{x}}_k,
    \label{eq:ts_model_new_output}
\end{equation}
where $\vec{h}_{k-1}(\vec{x}_k) = \left[h_{1,k-1}(\vec{x}_k)\quad\cdots\quad h_{C,k-1}(\vec{x}_k)\right]^\top\in\mathbb{R}^{C}$ is the vector of normalized activation degrees, and  $\vecsym{\hat{\Theta}}_{k-1}=\left[ \vecsym{\hat{\theta}}_{1,k-1}\quad\cdots\quad \vecsym{\hat{\theta}}_{C,k-1}\right]^\top\in\mathbb{R}^{n_x+1\times C}$ is the matrix of consequent coefficients, as estimated in the previous time step using \ac{rls}.

Given a number of rules, $C$, the \ac{ts} model creates a coarse fuzzy partitioning of the data space, and updates the parameters of first-order consequent functions to locally approximate the behavior of a system. An issue on online stream modeling concerns the lack of \textit{a priori} knowledge about the model structure, \ie, its number of rules \cite{kasabov2002denfis}. This is particularly relevant in fault prognostics since degradation dynamics are typically nonlinear, non-stationary, and different for each \ac{uut}.

\subsection{Uncertainty estimation}

Consider a state transition function given by a \ac{ts} model, with rules as in \eqref{eq:ith_fuzzy_rule}. The degradation propagation \eqref{eq:state_transition} can be rewritten as 
\begin{equation}
    \hat{x}_{k+N} = \vec{h}_{k}^\top\left(\vec{v}_{k+N,L}\right)\, 
    \vecsym{\hat{\Theta}}_{k}^\top\,\, \vec{\tilde{v}}_{k+N,L} + 
    \epsilon_{k+N}, \quad\forall N>0
    \label{eq:prog_ts_state_transition}
\end{equation}
where $\vec{h}_{k}(\cdot)$ and $\vecsym{\hat{\Theta}}_{k}$ are the normalized degrees of activation and consequent parameters for each rule with structure updated until time instant $k$; $N$ is the prediction horizon, and $\tilde{\vec{v}}_{k+N,L}$ is the augmented vector $\tilde{\vec{v}}_{k+N,L}\triangleq\left[1\quad\vec{v}_{k+N,L}^\top\right]^\top$. To account for prediction uncertainties, white Gaussian noise is added to \eqref{eq:prog_ts_state_transition} from
\begin{equation}
    \epsilon_{k}\sim\mathcal{N}\left(0,\sigma_{\epsilon}^2\right),
    \label{eq:prog_gn_var}
\end{equation}
where $\sigma_\epsilon^2$ is considered constant. The noise variance can be estimated through Monte Carlo simulations using the consequent parameters' covariance matrix estimated via \ac{rls} until time instant $k$ \cite{Camargos2020}. We provide a way to recursively track the covariance of estimation errors through the online learning operation, \ie, for time instances $n\in\mathbb{N}_{\leq k}$. The mean error is recursively tracked as
\begin{equation}
	\vecsym{\Delta}_{\epsilon,n} = \vec{\epsilon}_n - \vecsym{\hat{\mu}}_{\epsilon,n-1},
	\label{eq:error_mu_delta}
\end{equation}
\begin{equation}
	\vecsym{\hat{\mu}}_{\epsilon,n} = \vecsym{\hat{\mu}}_{\epsilon,n-1} + \frac{1}{n}\,\vecsym{\Delta}_{\epsilon,n},
	\label{eq:error_mu_update}
\end{equation}
in which $n$ is the total number of instances processed by the \ac{efs}. The initial mean error is $\vecsym{\hat{\mu}}_{\epsilon,0}=\vec{0}_{n_y\times 1}$ -- where $n_y=1$ in this case. Given the estimated mean error, the sum of squares is obtained recursively from
\begin{equation}
    \vec{M}_{\epsilon,n} = \vec{M}_{\epsilon,n-1} + (\vec{\epsilon}_n - \vecsym{\hat{\mu}}_{\epsilon,n-1})(\vec{\epsilon}_n - \vecsym{\hat{\mu}}_{\epsilon,n})^\top,
\end{equation}
being $\vec{M}_{\epsilon,0}=\vec{0}_{n_y\times n_y}$. The error covariance matrix at time instant $n$ is
\begin{equation}
    \vecsym{\Sigma}_{\epsilon,n} = \frac{\vec{M}_{\epsilon,n}}{n-1}.
\end{equation}
The variance $\sigma_\epsilon^2$ in \eqref{eq:prog_gn_var}, used for long-term prediction, is then approximated by
\begin{equation}
    \sigma_\epsilon^2 \approx \vecsym{\Sigma}_{\epsilon,k}.
\end{equation}

\subsection{Uncertainty propagation}
\label{sec:unc_prop}

After obtaining the initial uncertainty in one step estimates (Section 3.1), its long term propagation considers the input vector \eqref{eq:xkl_lags_nmdl} to be a vector composed of estimated random variables. Note that if $N=1$, then previous degradation states are known and, naturally, are non-random variables. Accordingly, the output $\hat{x}_{k+N}$ of the state transition relation \eqref{eq:prog_ts_state_transition} is also a random variable,
\begin{equation}
    \hat{x}_{k+N}^+ = \vec{h}_{k}^\top\left(\vec{z}_{k+N}\right)\, 
    \vecsym{\hat{\Theta}}_{k}^\top\,\, \vec{\tilde{v}}_{k+N,L}^+ + 
    \epsilon_{k+N}, \quad\forall N>0
    \label{eq:prog_ts_state_transition_rv}
\end{equation}
where $\vec{z}_{k+N}\triangleq \mathbb{E}[\vec{\tilde{v}}_{k+N,L}^+]$ are premise variables defined as the expected input vector. Computing variances in a multi-step prediction framework is needed for uncertainty propagation. The first step gives \begin{equation}
	\text{Var}\left(\hat{x}_{k+1}^+\right) =\text{Cov}\left( \vec{h}_{k}^\top\left(\vec{z}_{k+1}\right)\, 
	\vecsym{\hat{\Theta}}_{k}^\top\,\, \vec{\tilde{v}}_{k+1,L}^+\right) +
	\sigma_\epsilon^2.\label{eq:var_xkp1_1}
\end{equation}
As rule activation degrees, $\vec{h}_k(\cdot)$, are calculated based on the expected value of the random variable $\vec{\tilde{v}}_{k+1,L}^+$, then they can be considered constant; similar to the parameters vector. Let
\begin{align}
    \vecsym{\Xi}_N\triangleq\vec{h}_{k}^\top\left(\vec{z}_{k+N}\right)\, \vecsym{\hat{\Theta}}_{k}^\top. 
\end{align}
Thus Eq. \eqref{eq:var_xkp1_1} becomes:
\begin{align}
	\text{Var}\left(\hat{x}_{k+1}^+\right) &=
	\vecsym{\Xi}_1\,\,\text{Cov}\left(\vec{\tilde{v}}_{k+1,L}^+\right)\,\vecsym{\Xi}_1^\top
	 + 
	\sigma_\epsilon^2\nonumber\\
	&= \vecsym{\Xi}_1\,\,\vecsym{\Lambda}_{1}^L\,\,\vecsym{\Xi}_1^\top + 
	\sigma_\epsilon^2\nonumber\\
	&= \sigma_\epsilon^2\nonumber\\
	&= \lambda_1^2,
\end{align}
in which $\vecsym{\Lambda}_N^L\triangleq\mathrm{Cov}\left(\vec{\tilde{v}}_{k+N,L}^+\right)$, and $\lambda_N^2\triangleq\mathrm{Var}\left(\hat{x}_{k+N}^+\right)$. Note that $\vecsym{\Lambda}_{1}^L=0$, since previous degradation states are known at $N=1$. The $N$-step variance is computed recursively as 
\begin{equation}
    \text{Var}\left(\hat{x}_{k+N}^+\right) = \vecsym{\Xi}_N\,\,\vecsym{\Lambda}_{N}^L\,\,\vecsym{\Xi}_N^\top + 
    \sigma_\epsilon^2\label{eq:final_uncert}.
\end{equation}
The covariance matrix of the random vector $\vec{\tilde{v}}_{k+1,L}^+$ is
\begin{equation}
    \vecsym{\Lambda}_{N}^L=\begin{bmatrix}
        0 & 0 & \cdots & 0\\
        0 & \lambda_{N-1}^2 & \cdots & \lambda_{N-L}\lambda_{N-1}\hat{\rho}_{L,1}\\
        \vdots & \vdots & \ddots&\vdots\\
        0 & \lambda_{N-1}\lambda_{N-L}\hat{\rho}_{1,L}&\cdots&\lambda_{N-L}^2
    \end{bmatrix},
    \label{eq:cov_mat_unc_prop}
\end{equation}
where the first row and column contain zeros by default, due to matrix augmentation. Moreover, $\lambda_i^2=0$ when $i<0$, meaning that $x_{k+N}$ is known. The convariance matrix \eqref{eq:cov_mat_unc_prop} is weighted by Pearson correlation coefficients, $\hat{\rho}$, estimated by means of available \ac{uut} data.

Considering the degradation to be a random variable with Gaussian distribution, whose expected value is propagated by successive iterations of \eqref{eq:prog_ts_state_transition}, then \ac{rul} lower and upper bounds at an $(\alpha)$(100)\% significance level are given as
\begin{subequations}
    \begin{equation}
    \hat{r}_{\mathrm{lower},k} = \inf{\{N\in\mathbb{N}: \hat{x}_{k+N} + z_{1-\alpha/2}\,\nu_{N} 
    \geq\eta\}},
    \label{eq:rul_inf}
    \end{equation}
    \begin{equation}
    \hat{r}_{\mathrm{upper},k} = \inf{\{N\in\mathbb{N}: \hat{x}_{k+N} + z_{\alpha/2}\,\nu_{N} 
        \geq\eta\}}.
    \label{eq:rul_sup}
    \end{equation}
\end{subequations}

Representing, quantifying, forward propagating, and managing uncertainty are issues of utmost importance to support decision-making in practical engineering applications \cite{Sankararaman2015}. Nevertheless, there is a lack of effective uncertainty quantification approaches for multi-step prediction based on evolving fuzzy models. In this sense, the uncertainty quantification method described in this section, despite its relative simplicity, is an original contribution to evolving fuzzy modeling. The method enables fault prognostics in dynamic and time-varying environment.

\section{Experimental setup}
\label{sec:exp_set}

The case study reported in this section concerns the degradation of Li-ion batteries. This type of battery is found in industry and commercially, e.g., in electric vehicles, microgrids, and electronic devices \cite{Li2019jps,saha2009modeling}. The cycle aging datasets of four Li-ion batteries are provided by a testbed in the NASA Ames Prognostics Center of Excellence (PCoE). The testbed comprises commercial Li-ion 18650-sized rechargeable batteries from the Idaho National Laboratory; a programmable 4-channel DC electronic load and power supply; voltmeters, ammeters, and a thermocouple sensor suite; custom electrochemical impedance spectrometry equipment; and environmental chamber to impose different operational conditions. The batteries run at room temperature (23º C). Charging is done in constant mode at 1.5 A, until the voltage reaches 4.2 V. Discharging is performed at a constant current level of 2 A, until the battery voltage reaches 2.7 V \cite{saha2009modeling}.

\begin{figure}[h!]
	\centering
	\includegraphics[width=0.95\linewidth]{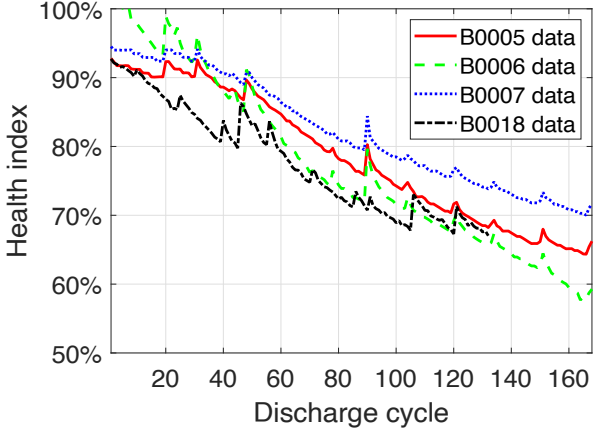}
	\caption{Percentage charge capacity along discharge cycles.}%
	\label{fig:rslt-hi}
\end{figure}

The health index (\ac{hi}) used in the experiments is the percentage charge capacity. When the batteries reach a 30\% deterioration in rated capacity (from 1.4 to 2 Ah), experiments are terminated \cite{saha2009modeling}. Therefore, the \ac{ft} is 70\%. \fgref{fig:rslt-hi} summarizes the datasets, namely, B0005, B0006, B0007, and B0018. The dataset B0006 is arbitrarily chosen as the training dataset. We compare three \ac{efs}s with each other and with a non-evolving method based on an \ac{arma} model.

\subsection{Parameter tuning}

Default hyperparameters were chosen for the different algorithms to be compared. For the \ac{ebets} approach, the choice is based on problem-agnostic recommendations \cite{Camargos2020}. The \ac{ebets} hyperparameters are $\omega_\textsc{\ac{ebets}}=95.45\%$, $\tau_\textsc{\ac{ebets}}=\ell+1$, $\gamma_\textsc{\ac{ebets}}=0.5$, and $\delta_\textsc{\ac{ebets}}=10^3$, with $\ell$ being the number of input lags or autoregressors. The \ac{xts} approach depends on the rule covariance initialization constant, $\Omega_{\textsc{\ac{xts}}}$, whose meaning is analogous to that of $\delta_\textsc{\ac{ebets}}$; therefore $\Omega_{\textsc{\ac{xts}}}=10^3$. The \ac{emg} approach uses as learning rate $\beta_\textsc{\ac{emg}}=0.05$; the unilateral confidence interval to define the \ac{emg} compatibility threshold is $\alpha_{\textsc{\ac{emg}}}=0.01$, the window size for the alert mechanism is $w_\textsc{\ac{emg}}=20$, and the initial dispersion matrix to create clusters is $\Sigma_\textsc{\ac{emg}}^{init}=10^{-3}\times \vec{I}_{\ell}$. 

The number of input lags $\ell$ is a free parameter to be optimized based on accuracy indices, namely, \ac{ra} and \ac{mape}. They are computed as follows \cite{saxena2008metrics}:
\begin{equation}
    \text{\ac{mape}}_k = \frac{100}{N} \sum_{i=k+1}^{k+N} 
    \left|\frac{x_i - \hat{x}_i}{x_i}\right|,
\end{equation}
\begin{equation}
    \text{\ac{ra}}_k = 1 - \frac{\left|r_k - \hat{r}_k\right|}{r_k},
\end{equation}
where $N$ is the number of forthcoming predictions until the \ac{uut} state reaches the threshold; $r_k$ and $\hat{r}_k$ are the actual and estimated \ac{rul} at $k$, respectively.

The training data, \ie, the data from B0006, and 20 samples of each test dataset, are used to validate a proper number of lags for each modeling approach. Let $I_k(.)$ be
\begin{align}
    I_k(\ell,\zeta,\varkappa) &= \acs{ra}_k(\ell,\zeta,\varkappa) + \left(1-\frac{\acs{mape}_k(\ell,\zeta,\varkappa)}{100}\right)\nonumber\\
    & \qquad + \left(1-\frac{\ell}{20}\right),
    \label{eq:index_bat}
\end{align}
where $\zeta\in\{\mathrm{B0006},\mathrm{B0007},\mathrm{B0018}\}$ is a testing battery;  $\varkappa\in\{\ac{ebets},\ac{xts},\ac{arma},\ac{emg}\}$ is an algorithm; and $\ell \in [1,~20]$ is the number of lags. \ac{arma} models consider $p$ and $q$ within $[1,~10]$. The number of lags arises as a result of the following maximization problem,
\begin{equation}
    \ell(\zeta,\varkappa) = \argmax_{l}\quad
    \frac{1}{4}\,\sum_{j\,\in\,\{5, 10, 15, 20\}} I_j(l,\zeta,\varkappa).
    \label{eq:index_bat_mean}
\end{equation}

\begin{figure}[ht]
	\centering
	\includegraphics[width=.95\linewidth]{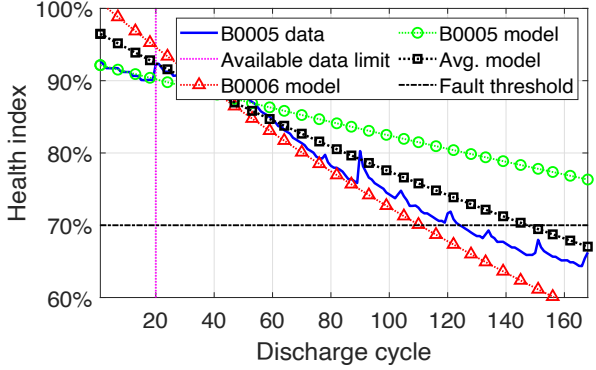}
	\caption{Combined exponential model using B0006 model and B0005 partial model.}%
	\label{fig:rslt_hyp_bparams}
\end{figure}

Index $I_k(.)$ \eqref{eq:index_bat} depends on the actual \ac{rul} of a testing battery to compute \ac{ra}$_k$. We propose an approximation for validation purpose based on a relation commonly used to quantify the charge capacity of Li-ion batteries,
\begin{equation}
	C(k;\vec{c}) = c_1\exp{c_2k} + c_3\exp{c_4k},
	\label{eq:bat_exp_mdl}
\end{equation}
in which the parameters' vector, $\vec{c}=[c_1\quad c_2\quad c_3\quad c_4]^\top$, is given using the known data from a battery and the least-squares method. The function \texttt{lsqcurvefit}\footnote{Available in \url{https://www.mathworks.com/help/optim/ug/lsqcurvefit.html}} is used to find the parameters' vector $\vec{c}$ in \eqref{eq:bat_exp_mdl} for the training battery B0006, which are used as a starting point to estimate the parameters of the same exponential model \eqref{eq:bat_exp_mdl} for the test batteries. \ac{rul} estimates for each test battery take the average between the model developed from the training data B0006 and the model found based on the first 20 test samples. Overall and average results are exemplified in \fgref{fig:rslt_hyp_bparams}. The same procedure is applied to batteries B0007 and B0018.

The parameters of the exponential relation \eqref{eq:bat_exp_mdl}, as portrayed by the data in \fgref{fig:rslt_hyp_bparams}, are listed in \tbref{tab:hyp_bat_params}. To find the parameters in \tbref{tab:hyp_bat_params} using B0006 training data, the least-squares starting point is $\vec{c}_0 = [1\quad1\quad1\quad0]^\top$. For the other batteries, the final coefficients for the B0006 model are used as starting point.

\begin{table}[ht]
    \caption{Least-squares fit of the exponential models' parameters based on battery charge capacity data.}
    \centering
    \label{tab:hyp_bat_params}
    \begin{tabular}{crrrr}
    \hline
    Battery & $c_1$ & $c_2$ & $c_3$ & $c_4$\\\hline
    B0006 & -0.4512 & 13.3905 & 1.0115 & 0.0033\\
    B0005 & -0.4512 & 13.3905 & 0.9226 & 0.0011\\
    B0007 & -0.4512 & 13.3905 & 0.9437 & 0.0010\\
    B0018 & -0.4512 & 13.3905 & 0.9305 & 0.0031\\\hline
    \end{tabular}
\end{table}

After defining the parameters of \eqref{eq:bat_exp_mdl} for each test dataset, \ac{rul} estimates are given and the index $I_k$ \eqref{eq:index_bat} is computed. With $I_j ~ \forall j$ in hands, we maximize \eqref{eq:index_bat_mean} to obtain the optimal number of lags. In particular, given the data $\zeta$ related to a battery, and an algorithm $\varkappa$, the following steps are performed to find the optimal number of lags:
\begin{enumerate}
    \item Find the parameters of an instance of \eqref{eq:bat_exp_mdl} using the training data, B0006; \label{po_bat_step1}
    \item Find the parameters of an instance of \eqref{eq:bat_exp_mdl} using the test data of battery $\zeta$, i.e., the first 20 samples of the respective dataset; \label{po_bat_step2}
    \item Provide a \acs{rul} estimate using the average of the predictions given by the models from Steps \ref{po_bat_step1} and \ref{po_bat_step2}; \label{po_bat_step3}
    \item For the different amounts of lags, from 2 to 20, manipulate the training data, B0006, using a Hankel matrix, accordingly; \label{po_bat_step4}
    \item Use algorithm $\varkappa$ to train a model for each amount of lags based on the data from Step \ref{po_bat_step4}; \label{po_bat_step5}
    \item Solve the maximation problem  \eqref{eq:index_bat_mean} to find the optimal number of lags, and optimal model.
\end{enumerate}

\section{Results and discussion}
\label{sec:res_dis}

The optimization problem \eqref{eq:index_bat_mean} defines the number of input lags for each algorithm-battery pair. The third column of \tbref{tab:battery_res} shows the number of lags chosen for each pair; the subsequent columns show the \ac{ra} for different starting prognostics points $t_P$. The symbol `*' indicates that the prognostics task was not carried out for the $t_P$. Additionally, `--' means the infeasibility of an algorithm to compute the \ac{rul} for the $t_P$. Infeasibility happens if long-term predictions converge to a value greater than the \ac{ft}, or have their slope changed to positive, thus never reaching the \ac{ft}. Notice that the lag column in \tbref{tab:battery_res} for \ac{arma} models corresponds to the parameter $p$, whereas the $q$ coefficient is zero for all cases, as found in validation. To compare algorithms fairly, the sample index ($s_i$) in which the prognostics start is set considering the $t_P$ and the number of input lags of each battery-algorithm pair. Using the \ac{ebets} algorithm as a basis, then $s_i=t_P-(\ell-3)$. For this reason, the B0018-\ac{xts} pair is unable to start the prognostics task at $t_P=20$, since $s_i=6$ is less than $\ell=17$.

\tbref{tab:battery_res} indicates that multivariate Gaussian models can better capture the information in some datasets. For instance, \ac{ebets} and \ac{emg} have shown similar results for battery B0005. However, this is not observed for batteries B0007 and B0018, in which non-evolving \ac{arma} models may eventually perform better. In general, different initial hyperparameters for the different algorithms may lead to slightly different results. In the absence of a fine-tuning procedure to set initial hyperparameters, evolving algorithms are more prone to develop sub-optimal models in the sense of long-term trends. Nevertheless, we highlight that \ac{ebets} hyperparameters come from a problem-agnostic methodology, \ie, from a method that does not require expert knowledge about the problem.

\begin{table}[ht]
    \caption{\acs{ra} for algorithm-battery pairs with prognostics starting at different $t_P$. Best values are in bold.}
    \centering
    \label{tab:battery_res}
    \small
    \begin{tabular}{ccrrrrrr}
    \hline
    \multirow{2}*{Bat.} & \multirow{2}*{Alg.} & \multirow{2}*{$\ell$} & \multicolumn{5}{c}{$t_P$}\\\cline{4-8}
    & & & 20 & 40 & 60 & 80 & 100\\\hline
    \multirowcell{4}{B0005\\fails at\\cycle\\125}
    & \acs{ebets} &  3  & \textbf{0.94} &          0.78 &          0.76 & \textbf{0.98} & \textbf{0.96}\\
    &   \acs{xts} &  9  &            -- &            -- &            -- &          0.95 &          0.91\\
    &  \acs{arma} &  1  &          0.77 &          0.83 &          0.86 &          0.74 &          0.82\\
    &   \acs{emg} &  5  &          0.89 & \textbf{0.98} & \textbf{0.94} &          0.91 & \textbf{0.96}\\\hline
    \multirowcell{4}{B0007\\fails at\\cycle\\166}
    & \acs{ebets} &  3  & \textbf{0.82} & \textbf{0.89} & \textbf{0.84} &          0.72 &          0.75\\
    &   \acs{xts} & 10  &          0.69 &          0.55 &            -- &            -- &          0.83\\
    &  \acs{arma} &  1  &          0.59 &          0.62 &          0.57 &          0.51 &          0.52\\
    &   \acs{emg} &  5  &          0.69 &          0.76 &          0.71 &          0.63 &            --\\\hline
    \multirowcell{4}{B0018\\fails at\\cycle\\97}
    & \acs{ebets} &  3  &          0.91 & \textbf{0.96} &          0.79 & \textbf{0.79} & *\\
    &   \acs{xts} & 17  &             * &          0.59 &            -- &            -- & *\\
    &  \acs{arma} &  1  &          0.80 &          0.78 & \textbf{0.91} &          0.57 & *\\
    &   \acs{emg} &  5  &          0.84 &          0.89 &            -- &            -- & *\\\hline
\end{tabular}
{\raggedright\small*\quad prognostics task not performed.\\--\quad algorithm's infeasibility to give the \acp{rul}.\par}
\end{table}

\begin{figure}[htb]
    \centering
    \includegraphics[width=.95\linewidth]{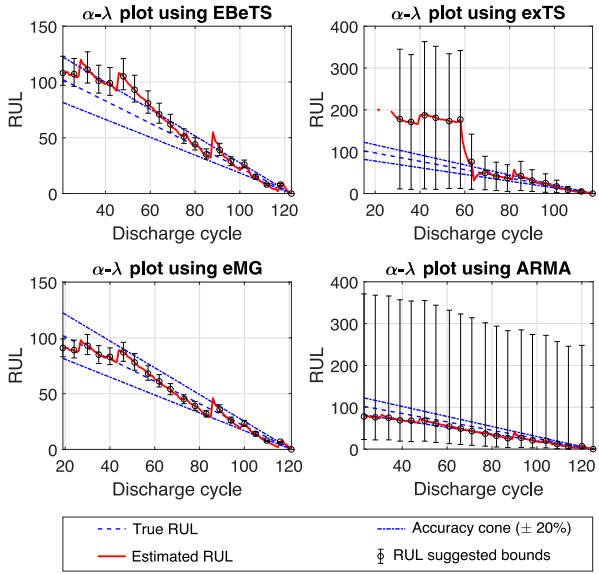}
    \caption{$\alpha-\lambda$ plot of the estimated \acs{rul} of battery B0005. The goal region is $\alpha=0.2$.}
    \label{fig:bat_rul_rmse_05}
\end{figure}

The $\alpha-\lambda$ plot for battery B0005 is shown in \fgref{fig:bat_rul_rmse_05}. The uncertainty is quantified for all evolving models using the online error tracking method with a 99\% confidence level. Uncertainty propagation within \ac{arma} models\footnote{A built-in function of the MATLAB System Identification Toolbox. Available in \url{https://www.mathworks.com/help/ident/ref/forecast.html}} yields too wide confidence intervals. Their bounds enclose the whole goal region, which is quite little useful to assist decision making. Similarities between \ac{ebets} and \ac{emg}, as noticed in \tbref{tab:battery_res}, is also perceived from \fgref{fig:bat_rul_rmse_05} for battery B0005. These methods provided the narrowest confidence intervals. In some experiments, the estimated \ac{rul} (red line) is missing. In these cases, the long-term prediction does not reach the \ac{ft}, as discussed previously.

\begin{figure}[htb]
    \centering
    \includegraphics[width=.95\linewidth]{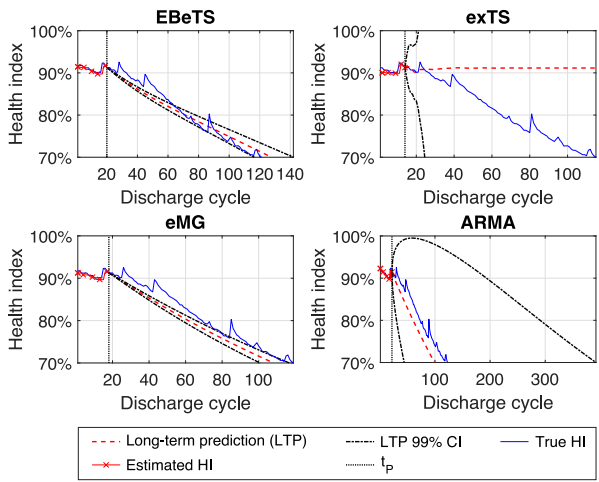}
    \caption{Long-term prediction, with 99\% confidence, of the different methods for Battery B0005.}
    \label{fig:rlst_bat_sd_b0005}
\end{figure}

\fgref{fig:rlst_bat_sd_b0005} shows the long-term prediction for battery B0005 and $t_P=20$, see dashed black line.
The dashed red line is the expected \ac{hi} propagated multiple steps ahead, while the dash-dotted black lines are its confidence intervals. In the \ac{arma} and \ac{xts} cases, such uncertainty interval becomes large enough to provide poor decision-making support, which is not the case for the remaining methods since they consider relationships among input features.

\newpage

\section{Conclusion}
\label{sec:conc}

\acp{efs} are promising methods to deal with nonlinear problems in non-stationary environments. Their structures are flexible, and their parameters can be updated recursively according to data stream changes. Structural learning from scratch, rapid recursive updates, and historical-data storage avoidance make \acp{efs} quite suitable to be used in real-time prognostics systems. We have shown the effectiveness of \acp{efs}, namely \ac{ebets} and \ac{emg}, in comparison to exTS and ARMA models, using a real-world benchmark dataset concerning the prognostics of charge capacity of Li-ion batteries. \acp{efs}-based models have offered online condition monitoring and a way of fusing multivariate data streams aiming at describing the multiple-stage battery-degradation phenomenon and providing prognostics. Furthermore, a framework to quantify and propagate uncertainties related to estimation errors has been improved to produce smooth confidence intervals. The proposed uncertainty quantification framework can be plugged into any \ac{efs} for real-time prognostics.

\begin{acknowledgment}
    This work was supported in part by the Brazilian agencies CNPq, FAPEMIG, CAPES, and in part by the PROPG-CAPES/FAPEAM Scholarship Program.
\end{acknowledgment}

\bibliographystyle{eusflat2021}
\bibliography{BIBeusflat2021}

%
%
%

\end{document}